\documentclass{wsmod-9x6}

\def\al{\alpha}
\def\be{\beta}
\def\ga{\gamma}
\def\de{\delta}
\def\ep{\epsilon}

\def\ka{\kappa}
\def\la{\lambda}

\def\si{\sigma}

\def\ps{\psi}
\def\cL{{\mathcal L}}

\def\etal{{\it et al.}}
\def\mn{{\mu\nu}}
\def\mnab{{\mu\nu\al\be}}
\def\tor#1#2#3{T^{#1}_{{\pt{#1}}#2#3}}
\def\pt#1{\phantom{#1}}
\def\fr#1#2{{{#1} \over {#2}}}
\def\half{{\textstyle{1\over 2}}}
\def\prt{\partial}
\def\bt{{\tilde b}}

\def\dt{{\tilde d}}
\def\gt{{\tilde g}}

\def\gev{\mbox{ GeV}}

\def\lrprt#1{\stackrel{\leftrightarrow}{\prt_{#1}}}

\def\gm#1#2#3{g^{(M)}_{#1#2#3}}

\def\glA#1{g^{(A)}_{#1}}
\def\gT#1{g^{(T)}_{#1}}

\def\tM#1#2#3{M_{#1#2#3}}
\def\xx#1#2{\xi^{(#1)}_{#2}}
\def\Tcoeff{(\xx 4 2 - 2 m \xx 5 8)}
\def\Acoeff{(\xx 4 4 - 2 m \xx 5 9)}
\def\Tcoef#1{(\xx 4 2 - 2 m_{#1} \xx 5 8)}
\def\Acoef#1{(\xx 4 4 - 2 m_{#1} \xx 5 9)}

\newcommand{\rf}[1]{(\ref{#1})}
\newcommand{\beq}{\begin{equation}}
\newcommand{\eeq}{\end{equation}}
\newcommand{\bea}{\begin{eqnarray}}
\newcommand{\eea}{\end{eqnarray}}

\begin{document}

\title{Lorentz Violation and Torsion}

\author{Neil Russell}

\address{Physics Department,
Northern Michigan University, \\
Marquette, MI 49855, U.S.A.}

\maketitle

\abstracts{
In this proceedings,
similarities between the structure
of theories with Lorentz violation and
theories with constant torsion in flat spacetime
are exploited to place bounds on
torsion components.
An example is given showing the analysis leading to
bounds on the axial-vector and mixed-symmetry components
of torsion,
based on a dual-maser experiment.
}

\section{Introduction}

In conventional Riemann-Cartan theory,
torsion is minimally coupled to a fermion.
Several nonminimal generalizations are possible,
including nonminimal couplings to a fermion,
nonminimal couplings involving a single particle of another species
(e.g., the photon),
and couplings involving more than one particle.
We limit consideration to signals of torsion effects
arising from nonminimal couplings of the torsion tensor to
one or more fermions only.
A more complete analysis is available
elsewhere.\cite{aknrTorsion}

We focus on the analysis of effects
relevant for laboratory experiments that
can be, or have been, done.
For simplicity,
torsion is taken as a
fixed background field in a Sun-centered inertial frame.
To consider fluctuations, the Nambu-Goldstone
and massive sectors\cite{NGmodes} have to be incorporated.
Other approaches, which include
ones seeking sensitivity to effects in
dynamical torsion theories,\cite{carrolfield,bsd}
are not considered here.
The literature on torsion includes
reviews by
Hehl \etal,\cite{hehl}
Shapiro,\cite{shapiro02} and
Hammond.\cite{hammond}
Possible bounds from Hughes-Drever experiments have been
discussed by L\"ammerzahl.\cite{lmrzhl}

Studies have shown that Lorentz violation
and the associated CPT violation
can arise in string theory.\cite{1989akss,1991akrp}
However,
it is possible to describe all the effects at
the level of effective field theory.\cite{1995akrp}
A systematic framework that encompasses
global Lorentz violation\cite{ck}
and local Lorentz violation\cite{akgrav} exists,
known as the Standard-Model Extension,
or SME.

The similarities between
lagrangian terms coupling fermions to torsion
and ones coupling fermions to Lorentz-violating backgrounds
means that
experimental sensitivity to  Lorentz-violation effects
cannot easily be decoupled from sensitivity to
torsion effects.
Here, we assume no Lorentz violation,
and interpret experimental results entirely in terms of torsion.

\section{Basics}
We adopt the conventions of Ref.\ \refcite{akgrav}.
The general metric $g_\mn$
has diagonal entries $(-1, 1,1,1)$
in the flat-space limit,
and
the antisymmetric tensor $\ep^\mnab$ is defined so that $\ep^{0123}=-1$.

The Riemann-Cartan curvature tensor,
denoted by $R_\mnab$,
consists of the usual Riemann curvature tensor
$\widetilde R_\mnab$ and added terms involving the contortion.
We are interested in the limit of spacetime with diagonal metric $(-1, 1,1,1)$,
in which case the Christoffel symbols are zero
and the usual curvature tensor $\widetilde R_\mnab$ vanishes.
We refer to this as `flat spacetime.'
In flat spacetime,
the Riemann-Cartan curvature tensor $R_\mnab$ does not necessarily vanish.

The torsion tensor $\tor\al\mu\nu$ is antisymmetric
in the second and third indices,
and so has 24 independent components.
We define the trace part $T^\mu$ and
the antisymmetric part $A^\mu$ of the torsion tensor as follows:
$T_\mu \equiv g^{\al\be} T_{\al\be\mu}$
and
$A^\nu \equiv \fr 1 6 \ep^{\al\be\mu\nu} T_{\al\be\mu}.$
The torsion tensor can be decomposed into irreducible components
\beq
T_{\mu\al\be}
=
\fr 1 3 (g_{\mu\al}T_\be - g_{\mu\be}T_\al)
- \ep_{\al\be\mn} A^\nu
+M_{\mu\al\be}
\, ,
\label{Tdecomp2}
\eeq
where $M_{\mu\al\be}$ is unique and is called the mixed-symmetry component.

\section{Fermions in flat spacetime with torsion}
The lagrangian for an electron of mass $m$ in flat spacetime
with all possible independent torsion couplings up to dimension five
is:\cite{aknrTorsion,akgrav}
\bea
\cL^{T,5}
&\approx&
\half i \bar \ps \ga^\mu \lrprt\mu \ps
- m \bar \ps \ps
\nonumber \\ &&
+ \xi^{(4)}_1 T_\mu \bar\ps\ga^\mu\ps
+ \xi^{(4)}_2 T_\mu \bar\ps\ga_5\ga^\mu\ps
+ \xi^{(4)}_3 A_\mu \bar\ps\ga^\mu\ps
+ \xi^{(4)}_4 A_\mu \bar\ps\ga_5\ga^\mu\ps
\nonumber \\ &&
+ \half i \xi^{(5)}_1 T^\mu \bar\ps\lrprt\mu \ps
+ \half \xi^{(5)}_2 T^\mu \bar\ps \ga_5 \lrprt\mu \ps
+ \half i \xi^{(5)}_3 A^\mu \bar\ps\lrprt\mu \ps
\nonumber \\ &&
+ \half \xi^{(5)}_4 A^\mu \bar\ps \ga_5\lrprt\mu\ps
+   \half i \xi^{(5)}_5 {M^\la}_{\mn} \bar\ps \lrprt\la \si^\mn\ps
+  \half i \xi^{(5)}_6 T_\mu \bar\ps \lrprt\nu \si^\mn\ps
\nonumber \\ &&
+ \half i \xi^{(5)}_7 A_\mu \bar\ps \lrprt\nu \si^\mn\ps
+ \half i \xi^{(5)}_8 \ep^{\la\ka\mn} T_\la \bar\ps \lrprt\ka \si_\mn\ps
\nonumber \\ &&
+ \half i \xi^{(5)}_9 \ep^{\la\ka\mn} A_\la \bar\ps \lrprt\ka \si_\mn\ps
\, .
\label{flatlagr}
\eea
This expression includes four coupling constants with dimension $m^0$:
$\xx 4 1$, \ldots, $\xx 4 4$,
and nine with dimension $m^{-1}$:
$\xx 5 1$, \ldots, $\xx 5 9$.
These terms can be arranged so as to match the Minkowski-spacetime limit
of the fermion-sector Lorentz-violating
lagrangian as given in Ref.\ \refcite{akgrav}.
In making this match,
we assume the torsion components are constants,
as are the SME coefficients.
Using Eqs. (12) to (14) of Ref.\ \refcite{akgrav},
and assuming zero torsion
and no electromagnetic field,
we have:
\bea
\cL^{SME}
&=&
\half i \bar \ps \ga^\mu \stackrel{\leftrightarrow}{\prt}_\mu \ps
- m \bar \ps \ps
\nonumber \\&&
- a_\mu \bar \ps \ga^\mu \ps
- b_\mu \bar\ps \ga_5 \ga^\mu \ps
\nonumber \\&&
- \half H_\mn \bar \ps \si^\mn \ps
- \half i c_\mn \bar \ps \ga^\mu \stackrel{\leftrightarrow}{\prt^\nu} \ps
- \half i d_\mn \bar \ps \ga_5 \ga^\mu \stackrel{\leftrightarrow}{\prt^\nu} \ps
\nonumber \\&&
- \half i e_\mu \bar \ps \stackrel{\leftrightarrow}{\prt^\mu} \ps
+ \half f_\mu \bar \ps \ga_5 \stackrel{\leftrightarrow}{\prt^\mu} \ps
\nonumber \\ &&
- \frac 1 4  i g_{\mn\la} \bar \ps \si^{\mn}
\stackrel{\leftrightarrow}{\prt^\la} \ps
\, .
\label{smeLagr2}
\eea
The SME coefficient $g_{\la\mn}$ appearing in the last term
can be decomposed in the same way as in Eq.\ \rf{Tdecomp2}:
\beq
{g_{\mu\nu}}^\la
=
\frac 1 3 (\gT\mu \de^\la_\nu - \gT\nu \de^\la_\mu)
- \ep_{\mn}^{\pt{\mn}\la\ka} g^{(A)}_\ka
+ g^{(M)\la}_\mn
\, ,
\label{gdecomp}
\eeq
where $\gT\mu$, $\glA \mu$ and $\gm \la\mu\nu$
are suitably-defined
trace,
axial-vector,
and mixed-symmetry components.
If we make this substitution and match
Eqs.\ \rf{flatlagr} and \rf{smeLagr2}
term by term,
a number of identities result, including, for example:
\bea
b_\mu  &=& - \xi^{(4)}_2 T_\mu - \xi^{(4)}_4 A_\mu \, ,
\label{beemu} \\
g^{(A)}_\ka &=& - 2(\xi^{(5)}_8 T_\ka + \xi^{(5)}_9 A_\ka) \, ,
\label{gAsyTor} \\
\gm \mu \nu \la &=& - 2 \xi^{(5)}_5 \tM \la\mu\nu \, .
\label{gMixTor}
\eea
A variety of experiments are sensitive to
Lorentz-violation coefficients,\cite{nraktables}
including $b_\mu$ and $g_{\la\mn}$,
and these equations show
they must also be sensitive to torsion effects.

\section{Connecting with experiments}
Experimental sensitivities in the case of ordinary matter
are to 40 tilde coefficients,
defined in Appendix B of Ref.\ \refcite{spaceprd}.
As an example,
we'll consider two of these,
$\bt_X$ and $\gt_{DX}$.
Using the decomposition in Eq.\ \rf{gdecomp},
they can be expressed
in terms of the irreducible
components of $g_{\la\mn}$:
\bea
\bt_X &=& b_X - m \glA X +  m \gm Y Z T \, , \\
\gt_{DX} &=& - b_X + m \glA X + 2 m \gm Y Z T \, .
\eea
If we now use Eqs.\ \rf{beemu}, \rf{gAsyTor}, and \rf{gMixTor},
we obtain relationships between
experimental tilde observables in the SME
and irreducible components of the torsion tensor:
\bea
\bt_X &=& -\Tcoeff T_X - \Acoeff A_X - 2m \xx 5 5 M_{T Y Z}
\label{tilde1}\, , \\
\gt_{DX} &=& +\Tcoeff T_X + \Acoeff A_X - 4m \xx 5 5 M_{T Y Z} \, .
\label{tilde2}
\eea
It follows that any experiment with sensitivity to
$\bt_X$ or $\gt_{DX}$
is also sensitive to the torsion components
$T_X$, $A_X$, and $M_{TYZ}$.
To investigate the expected sensitivity,
we next consider a specific experiment.

\section{Dual-maser experiment}
As an example,
consider the result
for a combination of coefficients\cite{kl}
given in the second line of Table II
of Ref.\ \refcite{cane},
which reports on
a recent He-Xe dual-maser experiment:
\beq
- \bt_X + 0.0034 \dt_X - 0.0034 \gt_{DX}
< (2.2 \pm 7.9) \times 10^{-32} \gev
\, .
\eeq
We note by inspection of Eqs.\ \rf{flatlagr} and \rf{smeLagr2}
that the $\dt_X$ coefficient is not relevant for torsion bounds,
and may therefore be taken as zero.
To introduce the experimental result,
we substitute Eqs.\ \rf{tilde1} and \rf{tilde2},
with the mass taken as that for the neutron,
$m_n \simeq 0.938 \gev$.
This gives one particular torsion combination bounded by this experiment.
Some simplification occurs
because the substitution
involves combinations of the same torsion terms,
and so we may neglect the third term since the factor of 0.0034 is small.
We extract a bound with a confidence level of about 90\%
by doubling the one-sigma uncertainty:
\bea
| \Tcoef n T_X +  \Acoef n A_X & +& 2 m_n \xx 5 5 \tM T Y Z |
\nonumber \\
&&< 1.6 \times 10^{-31}\gev \, .
\label{torbound}
\eea

The case of minimally-coupled torsion is recovered
when $\xx 4 4 = 3/4$ and all the other couplings are zero.
Then, Eq.\ \rf{torbound} yields the result
\beq
|A_X| < 2.1 \times 10^{-31} \gev \, .
\eeq
To extract additional torsion results,
we look at each term in Eq.\ \rf{torbound}
under the assumption that
the other terms vanish.
We find, for example, that
\beq
|\xx 5 5 \tM T Y Z | < 10^{-31} \, .
\eeq
Additional bounds on components of the torsion tensor
can be extracted by considering
the other terms in Eq.\ \rf{torbound},
and by considering the other bounds reported in this particular experiment.
More generally, the method adopted here can be used to
seek torsion bounds from other experiments.

% -----------------------------------------------------------


\begin{thebibliography}{xx}

\bibitem{aknrTorsion}
V.A.\ Kosteleck\'y, N.\ Russell and J.D.\ Tasson,
Phys.\ Rev.\ Lett.\ to appear,
[arXiv:0712.4393].

\bibitem{NGmodes}
R.\ Bluhm and V.A.\ Kosteleck\'y,
  Phys.\ Rev.\ D {\bf 71}, 065008 (2005);
R.\ Bluhm \etal,
  Phys.\ Rev.\ D, in press, [arXiv:0712.4119];
B.\ Altschul and V.A.\ Kosteleck\'y,
  Phys.\ Lett.\ B {\bf 628}, 106 (2005);
V.A.\ Kosteleck\'y and R.\ Potting,
  Gen.\ Rel.\ Grav.\ {\bf 37}, 1675 (2005).

\bibitem{carrolfield}
S.M.\ Carroll and G.B.\ Field,
Phys.\ Rev.\ {\bf D 50}, 3867 (1994).

\bibitem{bsd}
A.S.\ Belyaev, I.L.\ Shapiro, and M.A.B.\ do Vale,
Phys.\ Rev.\ {\bf D 75}, 034014 (2007).

\bibitem{hehl}
F.W.\ Hehl \etal,
Rev.\ Mod.\ Phys.\ {\bf 48}, 393 (1974).

\bibitem{shapiro02}
I.L.\ Shapiro,
Phys.\ Rep.\ {\bf 357}, 113 (2002).

\bibitem{hammond}
R.T.\ Hammond,
Rep.\ Prog.\ Phys.\ {\bf 65} 599 (2002).

\bibitem{lmrzhl}
C.\ L\"ammerzahl,
Phys.\ Lett.\ A {\bf 228}, 223 (1997).

\bibitem{1989akss}
V.A.\ Kosteleck\'y and S.\ Samuel,
  Phys.\ Rev.\ D {\bf 39}, 683 (1989);
  Phys.\ Rev.\ D {\bf 40}, 1886 (1989);
  Phys.\ Rev.\ Lett.\  {\bf 63}, 224 (1989).

\bibitem{1991akrp}
V.A.\ Kosteleck\'y and R.\ Potting,
Nucl.\ Phys.\ B {\bf 359}, 545 (1991).

\bibitem{1995akrp}
V.A.\ Kosteleck\'y and R.\ Potting,
Phys.\ Rev.\ D {\bf 51}, 3923 (1995).

\bibitem{ck}
D.\ Colladay and V.A.\ Kosteleck\'y,
  Phys.\ Rev.\ D {\bf 55}, 6760 (1997);
  Phys.\ Rev.\ D {\bf 58}, 116002 (1998);
V.A.\ Kosteleck\'y and R.\ Lehnert,
  Phys.\ Rev.\ D {\bf 63}, 065008 (2001).

\bibitem{akgrav}
V.A.\ Kosteleck\'y,
  Phys.\ Rev.\ D {\bf 69}, 105009 (2004).

\bibitem{nraktables}
For a tabulation, see
V.A.\ Kosteleck\'y and N.\ Russell,
arXiv:0801.0287.

\bibitem{spaceprd}
R.\ Bluhm \etal,
Phys.\ Rev.\ D {\bf 68}, 125008 (2003).

\bibitem{kl}
V.A.\ Kosteleck\'y and C.D.\ Lane,
Phys.\ Rev.\ D {\bf 60}, 116010 (1999).

\bibitem{cane}
F.\ Can\`e \etal,
Phys.\ Rev.\ Lett.\ {\bf 93}, 230801 (2004).

\end{thebibliography}
\end{document}